\documentstyle[amssymb,11pt]{article}

\newlength{\minitwocolumn}
\font\teneufm=eufm10
\font\seveneufm=eufm7
\font\fiveeufm=eufm5
\newfam\eufmfam
\textfont\eufmfam=\teneufm
\scriptfont\eufmfam=\seveneufm
\scriptscriptfont\eufmfam=\fiveeufm

\makeatletter
\@addtoreset{equation}{section}
\makeatother

\newtheorem{thm}{Theorem}[section]

\title{\bf
\Large{\bf
A REMARK ON GROUND STATE OF \\
BOUNDARY IZERGIN-KOREPIN MODEL}
}
\begin{document}
\maketitle
\begin{center}
{Takeo KOJIMA}
\\~\\
{\it
Department of Mathematics and Physics,
Graduate School of Science and Engineering,\\
Yamagata University, Jonan 4-3-16, Yonezawa 992-8510, Japan}\\
kojima@yz.yamagata-u.ac.jp
\end{center}
~\\

\begin{abstract}
We study
the ground state
of the boundary Izergin-Korepin model.
The boundary Izergin-Korepin model is
defined by so-called $R$-matrix and $K$-matrix for
$U_q(A_2^{(2)})$
which satisfy Yang-Baxter equation and
boundary Yang-Baxter equation.
The ground state associated with
identity $K$-matrix $\bar{K}(z)=id$ 
was constructed by W.-L.Yang and Y.-Z.Zhang in 
earlier study. 
We construct the free field realization of the ground state
associated with nontrivial diagonal $K$-matrix.
\end{abstract}

\section{Introduction}

There have been many developments in the field of exactly solvable models.
Various methods were invented to solve models.
The free field approach \cite{JM} provides a powerful method to study exactly solvable 
models. 
This paper is devoted to 
the free field approach to
boundary problem of exactly solvable statistical mechanics
\cite{JKKKM}.
Exactly solvable boundary model \cite{JKKKM, Sklyanin}
is defined by the solutions of the Yang-Baxter equation and
the boundary Yang-Baxter equation
$$
{K}_2(z_2)
{R}_{2,1}(z_1z_2)
{K}_1(z_1)
{R}_{1,2}(z_1/z_2)=
{R}_{2,1}(z_1/z_2)
{K}_1(z_1)
{R}_{1,2}(z_1z_2)
{K}_2(z_2).
$$
In this paper we are going to study the 
boundary Izergin-Korepin model 
defined by the solutions of 
the Yang-Baxter equation \cite{IK} and
the boundary Yang-Baxter equation \cite{BFKZ, MN} 
for the quantum group $U_q(A_2^{(2)})$.
We are going to diagonalize the infinite transfer matrix 
$T_\epsilon(z)$
of the boundary Izergin-Korepin model, by means of the free field approach.
For better understanding of the model that we are going to study,
we give comments on
the solutions
of the boundary Yang-Baxter equation.
The $R$-matrix 
associated with non-exceptional affine symmetry except for $D_n^{(2)}$ commute with each other
$$[\widehat{R}(z_1), \widehat{R}(z_2)]=0,$$
where $\widehat{R}(z)=P R(z)$
and $P(a\otimes b)=b \otimes a$.
Hence we know that
the identity $K$-matrix 
$$\bar{K}(z)=id$$
is a particular solution of the boundary Yang-Baxter equation
\cite{Sklyanin} for $A_n^{(1)}$, $B_n^{(1)}$, $C_n^{(1)}$,
$D_n^{(1)}$ and $A_n^{(2)}$ .
There exist more general solutions of the boundary Yang-Baxter equation.
The diagonal solutions of the boundary Yang-Baxter equation
for $A_n^{(1)}$, $B_n^{(1)}$, $C_n^{(1)}$,
$D_n^{(1)}$ and $A_n^{(2)}$ are classified
in \cite{BFKZ}.
For $A_n^{(1)}$
there exists the diagonal $K$-matrix that 
has one continuous free parameter.
However for $B_n^{(1)}$, $C_n^{(1)}$,
$D_n^{(1)}$ and $A_n^{(2)}$ there exist
only discrete solutions.
For example, for $A_2^{(2)}$
there exist three isolated solutions 
$\bar{K}(z)$.
\begin{eqnarray}
\bar{K}(z)=
\left(\begin{array}{ccc}
1&0&0\\
0&1&0\\
0&0&1
\end{array}\right),~~~
\left(\begin{array}{ccc}
z^2&0&0\\
0&\frac{\displaystyle
\pm \sqrt{-1} q^{\frac{3}{2}}+z}{
\displaystyle
\pm \sqrt{-1} q^{\frac{3}{2}}+z^{-1}}&0\\
0&0&1
\end{array}\right).\nonumber
\end{eqnarray}
In earlier study \cite{YZ} W.-L. Yang and Y.-Z. Zhang
constructed the free field realization of the ground state
associated with identity $K$-matrix
$\bar{K}(z)=id$ for  $A_2^{(2)}$.
In this paper
we construct the free field realization of the ground state
associated with nontrivial diagonal $K$-matrix
$\bar{K}(z)$ for  $A_2^{(2)}$.
This realization is the first example associated with
the discrete solutions of the boundary Yang-Baxter equation.
It is thought that the free field approach 
to the boundary problem
works for every discrete $K$-matrix of 
the affine symmetry.
Contrary to
$B_n^{(1)}$, $C_n^{(1)}$,
$D_n^{(1)}$ and $A_n^{(2)}$ case, 
there have been many papers on the free field
approach to boundary problem for 
$A_n^{(1)}$ symmetry.
For example, the higher-rank generalization \cite{FK, K}
and the elliptic deformation
\cite{MW, K} have been solved.

The plan of this paper is as follows.
In section 2 
we give physical picture of our problem.
We introduce the boundary Izergin-Korepin model
and introduce the transfer matrix 
$T_\epsilon(z)$.
In section 3 
we translate physical picture of our problem into
mathematical picture.
We construct the free field realization of
the  ground state $|B\rangle_\epsilon$
and give the diagonalization of 
the transfer matrix $T_\epsilon(z)$.

\section{Boundary Izergin-Korepin model}

In this section we formulate physical picture of
our problem.

\subsection{$R$-matrix and $K$-matrix}

In this section we introduce $R$-matrix $R(z)$ 
and $K$-matrix $K(z)$ associated with
the quantum group $U_q(A_2^{(2)})$.
We choose $q$ and $z$ such that
$-1<q^{\frac{1}{2}}<0$ and 
$|q^2|<|z|<|q^{-2}|$.
Let $\{v_+,v_0,v_-\}$ denotes the natural basis
of $V={\mathbb C}^3$.
We introduce the $R$-matrix 
${R}(z) \in {\rm End}(V \otimes V)$ \cite{IK}.
\begin{eqnarray}
{R}(z)=\frac{1}{\kappa(z)}
\left(\begin{array}{ccccccccc}
1&0 &0 &0 &0 &0 &0 &0 &0 \\
0 &b(z)&0 &c(z)&0 &0 &0 &0 &0 \\
 0&0 &d(z)&0 & e(z)&0 &f(z)&0 &0\\
 0&zc(z)&0 &b(z)&0 &0 &0 &0 &0 \\
 0& 0&-q^2z e(z)&0 &j(z)&0 &e(z)&0&0 \\
 0&0 &0 &0 &0 &b(z)&0 &c(z)&0 \\
 0&0 &n(z)&0 &-q^2z e(z)&0 &d(z)&0 &0\\
 0&0 &0 &0 &0 &zc(z)&0 &b(z)&0 \\
 0&0 &0 &0 &0 &0 &0 &0 &1
\end{array}
\right).
\end{eqnarray}
Here we have set the elements
\begin{eqnarray}
&&b(z)=\frac{q(z-1)}{q^2z-1},~c(z)=\frac{q^2-1}{q^2z-1},
\nonumber\\
&&d(z)=\frac{q^2(z-1)(qz+1)}{
(q^2z-1)(q^3z+1)},~e(z)=
\frac{q^{\frac{1}{2}}(z-1)(q^2-1)}{(q^2z-1)(q^3z+1)},
\nonumber\\
&&f(z)=\frac{(q^2-1)\{(q^3+q)z-(q-1)\}z}{
(q^2z-1)(q^3z+1)},~
n(z)=\frac{(q^2-1)\{(q^3-q^2)z+(q^2+1)\}}{
(q^2z-1)(q^3z+1)},\nonumber\\
&&j(z)=\frac{q^4z^2+(q^5-q^4-q^3+q^2+q-1)z-q}{
(q^2z-1)(q^3z+1)},\nonumber
\end{eqnarray}
and the normalizing function
\begin{eqnarray}
\kappa(z)=z\frac{
(q^6z;q^6)_\infty
(q^2/z;q^6)_\infty
(-q^5z;q^6)_\infty
(-q^3/z;q^6)_\infty}{
(q^6/z;q^6)_\infty
(q^2z;q^6)_\infty
(-q^5/z;q^6)_\infty
(-q^3z;q^6)_\infty}.
\end{eqnarray}
Here we have used the abbreviation
\begin{eqnarray}
(z;p)_\infty=\prod_{m=0}^\infty (1-p^mz).\nonumber
\end{eqnarray}
The matrix elements of ${R}(z)$
are given by
${R}(z)v_{j_1}\otimes v_{j_2}=
\sum_{k_1,k_2=\pm,0}v_{k_1}\otimes v_{k_2}
{R}(z)_{k_1,k_2}^{j_1,j_2}$,
where the ordering of the index is given by
$v_+\otimes v_+, v_+\otimes v_0, v_+\otimes v_-$,
$v_0\otimes v_+, v_0\otimes v_0, v_0\otimes v_-$,
$v_-\otimes v_+, v_-\otimes v_0, v_-\otimes v_-$.
The $R$-matrix ${R}(z)$ satisfies 
the Yang-Baxter equation
\begin{eqnarray}
{R}_{1,2}(z_1/z_2)
{R}_{1,3}(z_1/z_3)
{R}_{2,3}(z_2/z_3)=
{R}_{2,3}(z_2/z_3)
{R}_{1,3}(z_1/z_3)
{R}_{1,2}(z_1/z_2),\label{eqn:YBE}
\end{eqnarray}
the unitarity
\begin{eqnarray}
R_{1,2}(z_1/z_2)R_{2,1}(z_2/z_1)=id,
\end{eqnarray}
and the crossing symmetry
\begin{eqnarray}
R(z)_{k_1,k_2}^{j_1,j_2}=q^{\frac{j_2-k_2}{2}}
R(-q^{-3}z^{-1})_{-j_2,k_1}^{-k_2,j_1}.
\end{eqnarray}
We have set the normalizing function $\kappa(z)$
such that the minimal eigenvalue of the corner transfer
matrix becomes $1$ \cite{Bax, KKMMNN}.

We introduce $K$-matrix 
$K(z)\in {\rm End}(V)$
representing an interaction at the boundary,
which satisfies the boundary Yang-Baxter equation
\begin{eqnarray}
{K}_2(z_2)
{R}_{2,1}(z_1z_2)
{K}_1(z_1)
{R}_{1,2}(z_1/z_2)=
{R}_{2,1}(z_1/z_2)
{K}_1(z_1)
{R}_{1,2}(z_1z_2)
{K}_2(z_2).
\label{eqn:BYBE}
\end{eqnarray}
We consider only the diagonal solutions 
$K(z)={K}_\epsilon(z), (\epsilon=\pm,0)$
\cite{BFKZ, MN}.
\begin{eqnarray}
{K}_0(z)=\frac{\varphi_0(z)}{\varphi_0(z^{-1})}
\left(\begin{array}{ccc}
1&0&0\\
0&1&0\\
0&0&1
\end{array}\right),~~~~~
{K}_\pm (z)=\frac{\varphi_\pm(z)}{\varphi_\pm(z^{-1})}
\left(\begin{array}{ccc}
z^2&0&0\\
0&\frac{\displaystyle
\pm \sqrt{-1} q^{\frac{3}{2}}+z}{
\displaystyle
\pm \sqrt{-1} q^{\frac{3}{2}}+z^{-1}}&0\\
0&0&1
\end{array}\right).
\end{eqnarray}
Here we have set the normalizing function
\begin{eqnarray}
\varphi_\epsilon(z)=
\frac{\displaystyle
(q^8z;q^{12})_\infty 
(-q^9z^2;q^{12})_\infty}{
\displaystyle
(q^{12}z;q^{12})_\infty 
(-q^5z^2;q^{12})_\infty}
\times
\left\{
\begin{array}{cc}
1&~~~(\epsilon=0)\\
\frac{\displaystyle
(\pm \sqrt{-1} q^{\frac{9}{2}}z;q^6)_\infty 
(\mp \sqrt{-1} q^{\frac{7}{2}}z;q^6)_\infty}{
\displaystyle
(\pm \sqrt{-1} q^{\frac{1}{2}}z;q^6)_\infty 
(\mp \sqrt{-1} q^{\frac{3}{2}}z;q^6)_\infty}
&~~~(\epsilon=\pm)
\end{array}\right..
\label{def:varphi}
\end{eqnarray}
The matrix elements of 
${K}(z)$
are defined by
${K}(z)v_j=\sum_{k=\pm,0}
v_k{K}(z)_k^j$,
where the ordering of the index is given by
$v_+, v_0, v_-$.
The $K$-matrix $K(z) \in {\rm End}(V)$ 
satisfies the boundary unitarity
\begin{eqnarray}
K(z)K(z^{-1})=id,
\end{eqnarray}
and the boundary crossing symmetry
\begin{eqnarray}
K(z)_{k_1}^{k_2}
=\sum_{j_,j_2=\pm,0}
q^{\frac{1}{2}(k_1-j_1)}
R(-q^{-3} z^{-2})_{j_2,-j_1}^{-k_1,k_2}
K(-q^{-3}z^{-1})_{j_1}^{j_2}.
\end{eqnarray}
To put it another way 
we have chosen the normalizing function
$\varphi_\epsilon(z)$ such that 
the transfer matrix $T_\epsilon(z)$ 
(\ref{def:transfer})
acts on the ground state as $1$ (\ref{def:ground}).

\subsection{Physical picture}

In this section, following \cite{Sklyanin, JKKKM}, we introduce the
transfer matrix $T_\epsilon(z)$ that is generating function of
the Hamiltonian $H_\epsilon$ of our problem.
In order to consider physical problem,
it is convenient to introduce graphical interpretation
of $R$-matrix and $K$-matrix.
We present the $R$-matrix 
$R(z)_{k_1,k_2}^{j_1,j_2}$ in Fig.1.

\begin{center}
\unitlength 0.1in
\begin{picture}( 40.8500, 22.0000)(-12.8500,-25.1500)
%
\special{pn 8}%
\special{pa 2200 600}%
\special{pa 2200 1800}%
\special{fp}%
\special{sh 1}%
\special{pa 2200 1800}%
\special{pa 2220 1734}%
\special{pa 2200 1748}%
\special{pa 2180 1734}%
\special{pa 2200 1800}%
\special{fp}%
\special{pa 2800 1200}%
\special{pa 1600 1200}%
\special{fp}%
\special{sh 1}%
\special{pa 1600 1200}%
\special{pa 1668 1220}%
\special{pa 1654 1200}%
\special{pa 1668 1180}%
\special{pa 1600 1200}%
\special{fp}%
\put(30.0000,-12.0000){\makebox(0,0){$j_2$}}%
\put(22.0000,-4.0000){\makebox(0,0){$j_1$}}%
\put(22.0000,-20.0000){\makebox(0,0){$k_1$}}%
\put(22.0000,-22.0000){\makebox(0,0){$z_1$}}%
\put(14.0000,-12.0000){\makebox(0,0){$k_2$}}%
\put(10.0000,-12.0000){\makebox(0,0){$z_2$}}%
\put(16.0000,-26.0000){\makebox(0,0){Fig.1: $R$-matrix}}%
\put(2.0000,-12.0000){\makebox(0,0){$R(z_1/z_2)_{k_1,k_2}^{j_1,j_2}=$}}%
\end{picture}%
~~~~~~~~~~~~~~~~~~~~
\end{center}

~\\

We present the $K$-matrix 
$K(z)_{k}^{j}$ in Fig.2.

\begin{center}
\unitlength 0.1in
\begin{picture}( 28.9500, 24.0000)( -2.9500,-25.1500)
%
\special{pn 8}%
\special{pa 1200 400}%
\special{pa 2200 1200}%
\special{fp}%
\special{sh 1}%
\special{pa 2200 1200}%
\special{pa 2160 1144}%
\special{pa 2158 1168}%
\special{pa 2136 1174}%
\special{pa 2200 1200}%
\special{fp}%
\special{pa 2200 1200}%
\special{pa 1200 2000}%
\special{fp}%
\special{sh 1}%
\special{pa 1200 2000}%
\special{pa 1266 1974}%
\special{pa 1242 1968}%
\special{pa 1240 1944}%
\special{pa 1200 2000}%
\special{fp}%
%
\special{pn 8}%
\special{pa 2200 400}%
\special{pa 2200 2000}%
\special{fp}%
\special{pa 2600 400}%
\special{pa 2600 400}%
\special{fp}%
\special{pa 2600 400}%
\special{pa 2200 800}%
\special{fp}%
\special{pa 2600 600}%
\special{pa 2200 1000}%
\special{fp}%
\special{pa 2600 800}%
\special{pa 2200 1200}%
\special{fp}%
\special{pa 2600 1000}%
\special{pa 2200 1400}%
\special{fp}%
\special{pa 2600 1200}%
\special{pa 2200 1600}%
\special{fp}%
\special{pa 2600 1400}%
\special{pa 2200 1800}%
\special{fp}%
\special{pa 2600 1600}%
\special{pa 2200 2000}%
\special{fp}%
\special{pa 2600 1800}%
\special{pa 2400 2000}%
\special{fp}%
\special{pa 2400 400}%
\special{pa 2200 600}%
\special{fp}%
\put(10.0000,-2.0000){\makebox(0,0){$j$}}%
\put(10.0000,-22.0000){\makebox(0,0){$k$}}%
\put(10.0000,-18.0000){\makebox(0,0){$z^{-1}$}}%
\put(10.0000,-6.0000){\makebox(0,0){$z$}}%
\put(2.0000,-12.0000){\makebox(0,0){$K(z)_k^j=$}}%
\put(12.0000,-26.0000){\makebox(0,0){Fig.2: $K$-matrix}}%
\end{picture}%
\end{center}

~\\

Let us consider the half-infinite 
spin chain $\cdots \otimes V \otimes V \otimes V$.
Let us introduce the subspace ${\cal H}$
of the half-infinite 
spin chain by
\begin{eqnarray}
{\cal H}=Span \{\cdots \otimes
v_{p(N)} \otimes \cdots \otimes v_{p(2)} \otimes v_{p(1)}|
p(s)=0~~(s >>0)\}.
\end{eqnarray}
We introduce the vertex operator $\Phi_j(z),
(j=\pm,0)$ acting on
the space ${\cal H}$ by Fig.3.
To put it another way,
the vertex operator
$\Phi_j(z)$ is infinite-size matrix whose matrix elements
are given by
products of the $R$-matrix $R(z)_{k_1,k_2}^{j_1,j_2}$
\begin{eqnarray}
\left(\Phi_{j}(z)\right)
_{\cdots p(N)~ \cdots p(2)~ p(1)~}
^{\cdots p(N)' \cdots p(2)' p(1)'}
=\lim_{N \to \infty}
\sum_{\nu(1),\nu(2),\cdots,\nu(N)=\pm,0}
\prod_{j=1}^N
R(z)_{\nu(j-1), p(j)}^{\nu(j) p(j)'},
\end{eqnarray}
where $j=\nu(0)$.
In order to avoid divergence we restrict our consideration to the subspace 
${\cal H}$.
We introduce the dual vertex operator $\Phi_j^*(z^{-1}),
(j=\pm,0)$ 
acting on the space ${\cal H}$ by Fig.4.

\begin{center}
\unitlength 0.1in
\begin{picture}( 49.2000, 34.0000)( -2.5500,-35.1500)
%
\special{pn 8}%
\special{pa 1200 800}%
\special{pa 4600 800}%
\special{fp}%
\special{sh 1}%
\special{pa 4600 800}%
\special{pa 4534 780}%
\special{pa 4548 800}%
\special{pa 4534 820}%
\special{pa 4600 800}%
\special{fp}%
\special{pa 4200 400}%
\special{pa 4200 1200}%
\special{fp}%
\special{sh 1}%
\special{pa 4200 1200}%
\special{pa 4220 1134}%
\special{pa 4200 1148}%
\special{pa 4180 1134}%
\special{pa 4200 1200}%
\special{fp}%
\special{pa 3800 400}%
\special{pa 3800 1200}%
\special{fp}%
\special{sh 1}%
\special{pa 3800 1200}%
\special{pa 3820 1134}%
\special{pa 3800 1148}%
\special{pa 3780 1134}%
\special{pa 3800 1200}%
\special{fp}%
\special{pa 3400 400}%
\special{pa 3400 1200}%
\special{fp}%
\special{sh 1}%
\special{pa 3400 1200}%
\special{pa 3420 1134}%
\special{pa 3400 1148}%
\special{pa 3380 1134}%
\special{pa 3400 1200}%
\special{fp}%
\special{pa 3000 400}%
\special{pa 3000 1200}%
\special{fp}%
\special{sh 1}%
\special{pa 3000 1200}%
\special{pa 3020 1134}%
\special{pa 3000 1148}%
\special{pa 2980 1134}%
\special{pa 3000 1200}%
\special{fp}%
\put(48.0000,-8.0000){\makebox(0,0){$j$}}%
\put(14.0000,-6.0000){\makebox(0,0){$z$}}%
\put(42.0000,-2.0000){\makebox(0,0){$1$}}%
\put(24.0000,-6.0000){\makebox(0,0){$\cdots$}}%
\put(24.0000,-10.0000){\makebox(0,0){$\cdots$}}%
\put(6.0000,-8.0000){\makebox(0,0){$\Phi_j(z)=$}}%
\put(24.0000,-16.0000){\makebox(0,0){Fig.3: Vertex operator}}%
%
\special{pn 8}%
\special{pa 4600 2800}%
\special{pa 1200 2800}%
\special{fp}%
\special{sh 1}%
\special{pa 1200 2800}%
\special{pa 1268 2820}%
\special{pa 1254 2800}%
\special{pa 1268 2780}%
\special{pa 1200 2800}%
\special{fp}%
\special{pa 4200 2400}%
\special{pa 4200 3200}%
\special{fp}%
\special{sh 1}%
\special{pa 4200 3200}%
\special{pa 4220 3134}%
\special{pa 4200 3148}%
\special{pa 4180 3134}%
\special{pa 4200 3200}%
\special{fp}%
\special{pa 3800 2400}%
\special{pa 3800 3200}%
\special{fp}%
\special{sh 1}%
\special{pa 3800 3200}%
\special{pa 3820 3134}%
\special{pa 3800 3148}%
\special{pa 3780 3134}%
\special{pa 3800 3200}%
\special{fp}%
\special{pa 3400 2400}%
\special{pa 3400 3200}%
\special{fp}%
\special{sh 1}%
\special{pa 3400 3200}%
\special{pa 3420 3134}%
\special{pa 3400 3148}%
\special{pa 3380 3134}%
\special{pa 3400 3200}%
\special{fp}%
\special{pa 3000 2400}%
\special{pa 3000 3200}%
\special{fp}%
\special{sh 1}%
\special{pa 3000 3200}%
\special{pa 3020 3134}%
\special{pa 3000 3148}%
\special{pa 2980 3134}%
\special{pa 3000 3200}%
\special{fp}%
\put(48.0000,-28.0000){\makebox(0,0){$j$}}%
\put(42.0000,-22.0000){\makebox(0,0){$1$}}%
\put(24.0000,-26.2000){\makebox(0,0){$\cdots$}}%
\put(24.0000,-30.2000){\makebox(0,0){$\cdots$}}%
\put(14.0000,-26.2000){\makebox(0,0){$z^{-1}$}}%
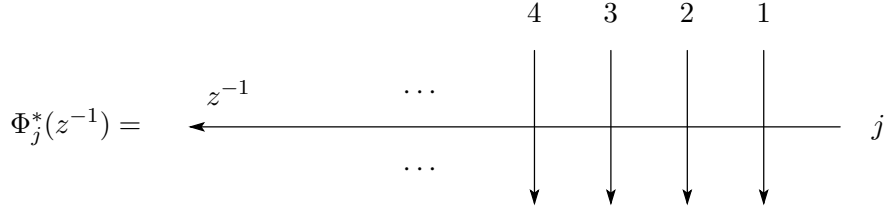
\put(24.0000,-36.0000){\makebox(0,0){Fig.4: Dual vertex operator}}%
\put(6.0000,-28.0000){\makebox(0,0){$\Phi_j^*(z^{-1})=$}}%
\put(38.0000,-2.0000){\makebox(0,0){$2$}}%
\put(34.0000,-2.0000){\makebox(0,0){$3$}}%
\put(30.0000,-2.0000){\makebox(0,0){$4$}}%
\put(38.0000,-22.0000){\makebox(0,0){$2$}}%
\put(34.0000,-22.0000){\makebox(0,0){$3$}}%
\put(30.0000,-22.0000){\makebox(0,0){$4$}}%
\end{picture}%
\end{center}

~\\

From the Yang-Baxter equation, we have the commutation relation
of the vertex operator $\Phi_j(z), (j=\pm,0)$
\begin{eqnarray}
\Phi_{j_2}(z_2)\Phi_{j_1}(z_1)=\sum_{k_1,k_2=\pm,0}
R(z_1/z_2)_{j_1,j_2}^{k_1,k_2}\Phi_{k_1}(z_1)\Phi_{k_2}(z_2).
\label{eqn:VOcom}
\end{eqnarray}
From the unitarity and the crossing symmetry,
we have the inversion relations
\begin{eqnarray}
g \Phi_{j_1}(z)\Phi_{j_2}^*(z)=\delta_{j_1,j_2}id,~~~
g \sum_{j=\pm,0}\Phi_j^*(z)\Phi_j(z)=id,
\label{eqn:VOinv}
\end{eqnarray}
where we have set 
$$g=\frac{1}{1+q}\frac{(q^2;q^6)_\infty (-q^3;q^6)_\infty}{
(q^6;q^6)_\infty (-q^5;q^6)_\infty}.
$$
From the crossing symmetry, we have
\begin{eqnarray}
\Phi_j^*(z)=q^{-\frac{j}{2}}\Phi_{-j}(-q^{-3}z).
\end{eqnarray}
We introduce the transfer matrix 
$\overline{T}_\epsilon(z)$
acting on the space ${\cal H}$ by Fig.5.
To put it another way
we introduce the transfer matrix 
$\overline{T}_\epsilon(z)$ by product of the vertex operators.
We define the "renormalized" transfer matrix 
$T_\epsilon(z)$ by
\begin{eqnarray}
T_\epsilon(z)=g \overline{T}_\epsilon(z)
=g \sum_{j,k=\pm,0}\Phi_j^*(z^{-1})K_\epsilon
(z)_j^k\Phi_k(z),~~~(\epsilon=\pm,0).
\label{def:transfer}
\end{eqnarray}

~\\

\begin{center}
\unitlength 0.1in
\begin{picture}
(58.1500,20.0000)(-0.1500,-21.1500)
%
\special{pn 8}%
\special{pa 400 400}%
\special{pa 400 400}%
\special{fp}%
\special{pa 800 600}%
\special{pa 4600 1000}%
\special{fp}%
\special{sh 1}%
\special{pa 4600 1000}%
\special{pa 4536 974}%
\special{pa 4548 994}%
\special{pa 4532 1014}%
\special{pa 4600 1000}%
\special{fp}%
\special{pa 4600 1000}%
\special{pa 800 1400}%
\special{fp}%
\special{sh 1}%
\special{pa 800 1400}%
\special{pa 868 1414}%
\special{pa 854 1394}%
\special{pa 864 1374}%
\special{pa 800 1400}%
\special{fp}%
\special{pa 4200 400}%
\special{pa 4200 1800}%
\special{fp}%
\special{sh 1}%
\special{pa 4200 1800}%
\special{pa 4220 1734}%
\special{pa 4200 1748}%
\special{pa 4180 1734}%
\special{pa 4200 1800}%
\special{fp}%
\special{pa 3800 400}%
\special{pa 3800 1800}%
\special{fp}%
\special{sh 1}%
\special{pa 3800 1800}%
\special{pa 3820 1734}%
\special{pa 3800 1748}%
\special{pa 3780 1734}%
\special{pa 3800 1800}%
\special{fp}%
\special{pa 3400 400}%
\special{pa 3400 1800}%
\special{fp}%
\special{sh 1}%
\special{pa 3400 1800}%
\special{pa 3420 1734}%
\special{pa 3400 1748}%
\special{pa 3380 1734}%
\special{pa 3400 1800}%
\special{fp}%
\special{pa 3000 400}%
\special{pa 3000 1800}%
\special{fp}%
\special{sh 1}%
\special{pa 3000 1800}%
\special{pa 3020 1734}%
\special{pa 3000 1748}%
\special{pa 2980 1734}%
\special{pa 3000 1800}%
\special{fp}%
\special{pa 2600 400}%
\special{pa 2600 1800}%
\special{fp}%
\special{sh 1}%
\special{pa 2600 1800}%
\special{pa 2620 1734}%
\special{pa 2600 1748}%
\special{pa 2580 1734}%
\special{pa 2600 1800}%
\special{fp}%
\special{pa 2200 400}%
\special{pa 2200 1800}%
\special{fp}%
\special{sh 1}%
\special{pa 2200 1800}%
\special{pa 2220 1734}%
\special{pa 2200 1748}%
\special{pa 2180 1734}%
\special{pa 2200 1800}%
\special{fp}%
%
\special{pn 8}%
\special{pa 4600 400}%
\special{pa 4600 1800}%
\special{fp}%
\special{pa 5000 400}%
\special{pa 4600 800}%
\special{fp}%
\special{pa 4800 400}%
\special{pa 4600 600}%
\special{fp}%
\special{pa 5000 600}%
\special{pa 4600 1000}%
\special{fp}%
\special{pa 5000 800}%
\special{pa 4600 1200}%
\special{fp}%
\special{pa 5000 1000}%
\special{pa 4600 1400}%
\special{fp}%
\special{pa 5000 1200}%
\special{pa 4600 1600}%
\special{fp}%
\special{pa 5000 1400}%
\special{pa 4600 1800}%
\special{fp}%
\special{pa 5000 1600}%
\special{pa 4800 1800}%
\special{fp}%
\put(42.0000,-2.0000){\makebox(0,0){1}}%
\put(10.0000,-4.0000){\makebox(0,0){$z$}}%
\put(10.0000,-16.0000){\makebox(0,0){$z^{-1}$}}%
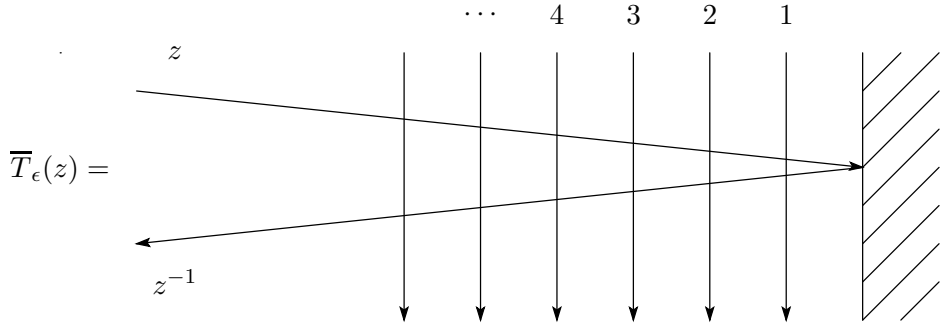
\put(28.0000,-22.0000){\makebox(0,0){Fig.5: Boundary transfer matrix}}%
\put(4.0000,-10.0000){\makebox(0,0){$\overline{T}_\epsilon(z)=$}}%
\put(38.0000,-2.0000){\makebox(0,0){$2$}}%
\put(34.0000,-2.0000){\makebox(0,0){$3$}}%
\put(30.0000,-2.0000){\makebox(0,0){$4$}}%
\put(26.0000,-2.0000){\makebox(0,0){$\cdots$}}%
\end{picture}%
\end{center}

~\\
From the commutation
relations of the vertex operators $\Phi_j(z),
\Phi_j^*(z)$ and the boundary Yang-Baxter equation 
(\ref{eqn:BYBE}),
we have the commutativity relation
of the transfer matrix $T_\epsilon(z)$, $(\epsilon=\pm,0)$.
\begin{eqnarray}
~[T_\epsilon(z_1),T_\epsilon(z_2)]=0,~~~~({\rm for~any}~z_1,z_2).
\end{eqnarray}
The commutativity of the transfer matrix
ensures that, 
if the transfer matrices $T_\epsilon(z)$ are
diagonalizable, the transfer matrices 
$T_\epsilon(z)$ are
diagonalized by the basis that is independent of
the spectral parameter $z$.
From the unitarity and the crossing symmetry,
we have
\begin{eqnarray}
T_\epsilon(z)T_\epsilon(z^{-1})=id,~~~
T_\epsilon(-q^{-3}z^{-1})=T_\epsilon(z).
\end{eqnarray}
The Hamiltonian $H_{IK}$ of the boundary
Izergin-Korepin model
is obtained by
\begin{eqnarray}
\frac{-1}{4(q-q^{-1})(q^{\frac{3}{2}}+q^{-\frac{3}{2}})}~
H_{IK}
=\left.\frac{d}{dz}T_\epsilon(z)\right|_{z=1}+const.
\end{eqnarray}
The Hamiltonian $H_{IK}$
is written as
\begin{eqnarray}
H_{IK}=H_1^b+\sum_{j=1}^\infty
H_{j+1,j},
\end{eqnarray}
where we have set 
$H^b \in {\rm End}(V)$ and
$H \in {\rm End}(V \otimes V)$ by
\begin{eqnarray}
H^b&=&
\left\{\begin{array}{cc}
0,&~~~(\epsilon=0)\\
4(q-q^{-1})\{-(q^{\frac{3}{2}}+q^{-\frac{3}{2}})\lambda_3
-\frac{\sqrt{3}}{3}(q^{\frac{3}{2}}-q^{-\frac{3}{2}}
\pm 2\sqrt{-1}) \lambda_8\},
&~~~(\epsilon=\pm)
\end{array}
\right.,
\end{eqnarray}
\begin{eqnarray}
H&=&
(q^{\frac{1}{2}}+q^{-\frac{1}{2}})(q^2+q^{-2})
(\lambda_1 \otimes \lambda_1+\lambda_2 \otimes \lambda_2)
\nonumber\\
&+&
(q^{\frac{1}{2}}+q^{-\frac{1}{2}})(q^2-q^{-2})\sqrt{-1}
(-\lambda_1\otimes \lambda_2
+\lambda_2\otimes \lambda_1)\nonumber\\
&+&
2(q^{\frac{1}{2}}+q^{-\frac{1}{2}})
\lambda_3 \otimes \lambda_3\nonumber\\
&+&
(q^{\frac{3}{2}}+q^{-\frac{3}{2}})(q+q^{-1})
(\lambda_4 \otimes \lambda_4+\lambda_5 \otimes \lambda_5
+\lambda_6 \otimes \lambda_6+\lambda_7 \otimes \lambda_7)
\nonumber\\
&+&
(q^{\frac{3}{2}}+q^{-\frac{3}{2}})(q-q^{-1})\sqrt{-1}
(\lambda_4 \otimes \lambda_5-\lambda_5 \otimes \lambda_4
+\lambda_6 \otimes \lambda_7-\lambda_7 \otimes \lambda_6)
\nonumber\\
&+&
(q-q^{-1})^2
(\lambda_4 \otimes \lambda_6+\lambda_6 \otimes \lambda_4
-\lambda_5 \otimes \lambda_7-\lambda_7 \otimes \lambda_5)
\nonumber\\
&+&
(q^2-q^{-2})\sqrt{-1}
(-\lambda_4 \otimes \lambda_7+\lambda_7 \otimes \lambda_4
-\lambda_5 \otimes \lambda_6+\lambda_6 \otimes \lambda_5)
\nonumber\\
&+&
\frac{2}{3}(-(q^{\frac{1}{2}}+q^{-\frac{1}{2}})
+2(q^{\frac{3}{2}}+q^{-\frac{3}{2}})+2(q^{\frac{5}{2}}+
q^{-\frac{5}{2}}))
\lambda_8 \otimes \lambda_8
\nonumber\\
&+&\frac{1}{3\sqrt{3}}
(-(q^{\frac{1}{2}}+q^{-\frac{1}{2}})+2
(q^{\frac{3}{2}}+q^{-\frac{3}{2}})-(q^{\frac{5}{2}}+q^{-\frac{5}{2}}))(
\lambda_8 \otimes id+
id \otimes \lambda_8).
\end{eqnarray}
Here we have used Gell-Mann matrices 
$\lambda_1,\lambda_2,\cdots,\lambda_8$
satisfying ${\rm Tr}(\lambda_j \lambda_k)=2\delta_{j,k}$.
\begin{eqnarray}
&&
\lambda_1=\left(\begin{array}{ccc}
0&0&1\\
0&0&0\\
1&0&0
\end{array}\right),
~
\lambda_2=\left(\begin{array}{ccc}
0&0&\sqrt{-1}\\
0&0&0\\
-\sqrt{-1}&0&0
\end{array}\right),
~
\lambda_3=\left(\begin{array}{ccc}
1&0&0\\
0&0&0\\
0&0&-1
\end{array}\right),\nonumber
\\
&&
\lambda_4=\left(\begin{array}{ccc}
0&1&0\\
1&0&0\\
0&0&0
\end{array}\right),~
\lambda_5=
\left(\begin{array}{ccc}
0&-\sqrt{-1}&0\\
\sqrt{-1}&0&0\\
0&0&0
\end{array}\right),
~
\lambda_6=\left(\begin{array}{ccc}
0&0&0\\
0&0&1\\
0&1&0
\end{array}\right),\nonumber\\
&&
\lambda_7=\left(\begin{array}{ccc}
0&0&0\\
0&0&\sqrt{-1}\\
0&-\sqrt{-1}&0
\end{array}\right),
~~~
\lambda_8=
\frac{1}{\sqrt{3}}
\left(\begin{array}{ccc}
1&0&0\\
0&-2&0\\
0&0&1
\end{array}\right).\nonumber
\end{eqnarray}
The diagonalization of the Hamiltonian $H_{IK}$
is reduced to those of the transfer matrix $T_\epsilon(z)$.
Let us consider the eigenvector problem
\begin{eqnarray}
T_\epsilon(z)|v\rangle=
t_\epsilon(z)|v\rangle.
\end{eqnarray}
We have chosen the normalizing function
$\varphi_\epsilon(z)$ (\ref{def:varphi}) such that 
the ground state $|B\rangle_\epsilon$
satisfies
\begin{eqnarray}
T_\epsilon(z)|B \rangle_\epsilon=|B \rangle_\epsilon,
~~~(\epsilon=\pm,0).
\label{def:ground}
\end{eqnarray}
We would like to construct the ground state 
$|B\rangle_\epsilon$ and would like to diagonalize the Hamiltonian $H_{IK}$.

\section{Free field realization}

In this section 
we give mathematical formulation of our problem.
We construct the free field realization
of the ground state $|B\rangle_\epsilon$, and give
the diagonalization of 
the transfer matrix $T_\epsilon(z)$.

\subsection{Mathematical picture}

In order to diagonalize 
the transfer matrix $T_\epsilon(z)$,
we follows the strategy proposed in \cite{JM, JKKKM}.
The corner transfer matrix method \cite{Bax, KKMMNN}
suggests
that we identify 
${\cal H}$ with the integrable highest-weight
representation $V(\Lambda_1)$ of the quantum group $U_q(A_2^{(2)})$,
because the characters of 
${\cal H}$ and $V(\Lambda_1)$ coincide.
We note that the representation
$V(\Lambda_1)$ is the only one level-1 
integrable highest-weight representation
of $U_q(A_2^{(2)})$.
We identify the line $\Phi_j(z)$ in Fig.3
with the components 
$\widetilde{\Phi}_j(z)$ of the vertex operator 
$\widetilde{\Phi}(z)$, and
the line $\Phi_j^*(z)$ in Fig.4 
with the components 
$\widetilde{\Phi}_j^*(z)$ of the dual vertex operator 
$\widetilde{\Phi}^*(z)$.
\begin{eqnarray}
&&\widetilde{\Phi}(z) : V(\Lambda_1) 
\rightarrow V(\Lambda_1)\otimes V_z,~~~
\widetilde{\Phi}(z)=\sum_{j=\pm,0}
\widetilde{\Phi}_j(z)\otimes v_j,
\label{def:VOI}
\\
&&\widetilde{\Phi}^*(z) :
V(\Lambda_1)\otimes V_z \rightarrow V(\Lambda_1),~~~
\widetilde{\Phi}_j^*(z)|v \rangle
=\widetilde{\Phi}^*(z)(|v\rangle \otimes v_j).
\end{eqnarray}
Here $V_z$ is the evaluation representation.
The vertex operator $\widetilde{\Phi}_j(z)$ for 
$U_q(A_2^{(2)})$
satisfies exactly the same functional relations
as those of $\Phi_j(z)$.
\begin{eqnarray}
&&\widetilde{\Phi}_{j_2}(z_2)
\widetilde{\Phi}_{j_1}(z_1)=\sum_{k_1,k_2=\pm,0}
R(z_1/z_2)_{j_1,j_2}^{k_1,k_2}
\widetilde{\Phi}_{k_1}(z_1)
\widetilde{\Phi}_{k_2}(z_2),\\
&&
g \widetilde{\Phi}_{j_1}(z)\widetilde{\Phi}_{j_2}^*(z)
=\delta_{j_1,j_2}id,~~~
\widetilde{\Phi}_j^*(z)=q^{-\frac{j}{2}}
\widetilde{\Phi}_{-j}(-q^{-3}z).
\end{eqnarray}
Let us set the transfer matrix 
$\widetilde{T}_\epsilon(z), (\epsilon=\pm,0)$ by
\begin{eqnarray}
\widetilde{T}_\epsilon(z)=
g \sum_{j,k=\pm,0}
\widetilde{\Phi}_j^*(z^{-1})
K_\epsilon(z)_j^k \widetilde{\Phi}_k(z),~~~(\epsilon=\pm,0).
\end{eqnarray}
We have the commutativity of the transfer matrix,
the unitarity and the crossing symmetry.
\begin{eqnarray}
&&~[\widetilde{T}_\epsilon(z_1),
\widetilde{T}_\epsilon(z_2)]=0,~~~~({\rm for~any}~z_1,z_2),\\
&&
\widetilde{T}_\epsilon(z)
\widetilde{T}_\epsilon(z^{-1})=id,
~~~
\widetilde{T}_\epsilon(-q^{-3}z^{-1})=
\widetilde{T}_\epsilon(z).
\end{eqnarray}
Let us set the ground state 
$|\widetilde{B}\rangle_\epsilon \in V(\Lambda_1)$ by
\begin{eqnarray}
\widetilde{T}_\epsilon(z)|\widetilde{B}\rangle_\epsilon=
|\widetilde{B}\rangle_\epsilon,~~~(\epsilon=\pm,0).
\end{eqnarray}
Following 
the strategy proposed in \cite{JM, JKKKM},
we consider our problem upon the
following identification.
\begin{eqnarray}
{\cal H}=V(\Lambda_1),~~\Phi_j(z)=\widetilde{\Phi}_j(z),
~~\Phi_j^*(z)=\widetilde{\Phi}_j^*(z),~~
T_\epsilon(z)=
\widetilde{T}_\epsilon(z),~~|B\rangle_\epsilon=
|\widetilde{B}\rangle_\epsilon.
\end{eqnarray}
In order to study the excitations
we introduce type-II vertex operator 
$\widetilde{\Psi}_\mu^*(z), (\mu=\pm,0)$.
\begin{eqnarray}
&&\widetilde{\Psi}^*(z) :
V_z \otimes V(\Lambda_1) \rightarrow V(\Lambda_1),~~~
\widetilde{\Psi}_\mu^*(z)|v \rangle
=\widetilde{\Psi}^*(z)(v_\mu \otimes |v\rangle ).
\label{def:VOII}
\end{eqnarray}
Type-II vertex operator 
$\widetilde{\Psi}_\mu^*(\xi)$
satisfies
\begin{eqnarray}
\widetilde{\Phi}_j(z)\widetilde{\Psi}_\mu^*(\xi)=
\tau(z/\xi)
\widetilde{\Psi}_\mu^*(\xi)\widetilde{\Phi}_j(z),~~~
(j, \mu=\pm,0).
\label{eqn:VOI-II}
\end{eqnarray}
Here we have set
\begin{eqnarray}
\tau(z)=z^{-1}
\frac{\Theta_{q^6}(q^5z)\Theta_{q^6}(-q^4z)}{
\Theta_{q^6}(q^5z^{-1})\Theta_{q^6}(-q^4z^{-1})},
~~~\Theta_p(z)=(p;p)_\infty(z;p)_\infty (pz^{-1};p)_\infty.
\end{eqnarray}
Let us set the vectors $|\xi_1,\xi_2,\cdots,\xi_N
\rangle_{\mu_1,\mu_2,\cdots,\mu_N,\epsilon} 
\in V(\Lambda_1)$, 
$(\mu_1,\mu_2,\cdots,\mu_N,\epsilon=\pm,0)$ by
\begin{eqnarray}
|\xi_1,\xi_2,\cdots,\xi_N
\rangle_{\mu_1,\mu_2,\cdots,\mu_N,\epsilon}
=\widetilde{\Psi}_{\mu_1}^*(\xi_1)
\widetilde{\Psi}_{\mu_2}^*(\xi_2)
\cdots 
\widetilde{\Psi}_{\mu_N}^*(\xi_N)|B\rangle_\epsilon.
\end{eqnarray}
From the commutation relation (\ref{eqn:VOI-II}) we have
\begin{eqnarray}
&&\widetilde{T}_\epsilon(z)
|\xi_1,\xi_2,\cdots,\xi_N
\rangle_{\mu_1,\mu_2,\cdots,\mu_N,\epsilon}\nonumber\\
&=&
\prod_{j=1}^N 
\tau(z/\xi_j)\tau(-1/q^3z\xi_j)~
|\xi_1,\xi_2,\cdots,\xi_N
\rangle_{\mu_1,\mu_2,\cdots,\mu_N,\epsilon}.
\end{eqnarray}
Comparing with the Bethe ansatz calculation
\cite{VR}
we conclude that
the vectors 
$|\xi_1,\xi_2,\cdots,\xi_N
\rangle_{\mu_1,\mu_2,\cdots,\mu_N,\epsilon}$ are the basis of
the space of state of the Izergin-Korepin model.
In order to construct 
the ground state 
$|B\rangle_\epsilon \in V(\Lambda_1)$,
it is convenient to
introduce the free field realization.

\subsection{Vertex operator}

In this section we give the free field realization
of the vertex operator $\widetilde{\Phi}_j(z), (j=\pm,0)$
\cite{Matsuno, JingMisra, HYZ}.
Let us introduce the bosons 
$a_m, (m \in {\mathbb Z}_{\neq 0})$ as following
\cite{Matsuno, Jing, JingMisra, HYZ}.
\begin{eqnarray}
~[a_m,a_n]=\delta_{m+n}\frac{[m]_q}{m}([2m]_q-(-1)^m[m]_q).
\end{eqnarray}
Here we have used $q$-integer
\begin{eqnarray}
[n]_q=\frac{q^n-q^{-n}}{q-q^{-1}}.
\end{eqnarray}
Let us set the zero-mode operators $P,Q$ by
\begin{eqnarray}
~[a_m,P]=[a_m,Q]=0,~[P,Q]=1.
\end{eqnarray}
The integrable highest weight representation 
$V(\Lambda_1)$
of $U_q(A_2^{(2)})$ is realized by
\begin{eqnarray}
V(\Lambda_1)={\mathbb C}[a_{-1},a_{-2},\cdots]
\oplus_{n \in {\mathbb Z}}e^{nQ}|\Lambda_1 \rangle,~~
|\Lambda_1\rangle=e^{\frac{Q}{2}}|0\rangle.
\end{eqnarray}
The vacuum vector $|0\rangle$ is characterized by
\begin{eqnarray}
a_m|0\rangle=0,~(m>0),~~P|0\rangle=0. 
\end{eqnarray}
We give the free field realizations of
the vertex operators.
Let us set 
$\epsilon(q)=
([2]_{q^{\frac{1}{2}}})^{\frac{1}{2}}$.
The highest elements of the vertex operators are given by
\begin{eqnarray}
\widetilde{\Phi}_-(z)&=&\frac{1}{\epsilon(q)}
e^{P(z)}e^{Q(z)}e^Q(-zq^4)^{P+\frac{1}{2}},\\
\widetilde{\Psi}_+^*(z)&=&\frac{1}{\epsilon(q)}
e^{P^*(z)}e^{Q^*(z)}e^{-Q}(-qz)^{-P+\frac{1}{2}}.
\end{eqnarray}
Here we have set the auxiliary operators 
$P(z),Q(z),P^*(z),Q^*(z)$ by
\begin{eqnarray}
P(z)&=&\sum_{m>0}\frac{a_{-m}}{[2m]_q-(-1)^m[m]_q}
q^{\frac{9m}{2}}z^m,\\
Q(z)&=&-\sum_{m>0}\frac{a_m}{[2m]_q-(-1)^m[m]_q}
q^{-\frac{7m}{2}}z^{-m},
\\
P^*(z)&=&
-\sum_{m>0}\frac{a_{-m}}{[2m]_q-(-1)^m[m]_q}
q^{\frac{m}{2}}z^m,\\
Q^*(z)&=&
\sum_{m>0}\frac{a_m}{[2m]_q-(-1)^m[m]_q}
q^{-\frac{3m}{2}}z^{-m}.
\end{eqnarray}
The other elements of the vertex operators are given by
the intertwining relations 
(\ref{def:VOI}) and (\ref{def:VOII}).
\begin{eqnarray}
\widetilde{\Phi}_0(z)&=&
\widetilde{\Phi}_-(z)x_0^--qx_0^- \widetilde{\Phi}_-(z),\\
\widetilde{\Phi}_+(z)&=&q^{\frac{1}{2}}
(\widetilde{\Phi}_0(z)x_0^--x_0^-\widetilde{\Phi}_0(z)),\\
\widetilde{\Psi}_0^*(z)&=&
x_0^+ \widetilde{\Psi}_+^*(z)-q^{-1}
\widetilde{\Psi}_+^*(z)x_0^+,\\
\widetilde{\Psi}_-^*(z)&=&
q^{-\frac{1}{2}}
(x_0^+ \widetilde{\Psi}_0^*(z)-\widetilde{\Psi}_0^*(z)x_0^+).
\end{eqnarray}
The elements $x_m^\pm, (m \in {\mathbb Z})$ 
are given by integral of the currents
$x^\pm(w)=\sum_{m \in {\mathbb Z}}x_m^\pm w^{-m}$,
\begin{eqnarray}
x_m^\pm=\oint \frac{dw}{2\pi \sqrt{-1}}w^{m-1}x^\pm(w),~~~
(m \in {\mathbb Z}).
\end{eqnarray}
Here the free field realizations
of the currents $x^\pm(w)$ are given by
\begin{eqnarray}
x^\pm(w)=\epsilon(q)e^{R^\pm(w)}e^{S^\pm(w)}
e^{\pm Q}w^{\pm P+\frac{1}{2}},
\end{eqnarray}
where we have set the auxiliary operators 
$R^\pm(w), S^\pm(w)$ by
\begin{eqnarray}
R^\pm(w)&=&\pm \sum_{m>0}\frac{a_{-m}}{[m]_q}
q^{\mp \frac{m}{2}}w^m,\\
S^\pm(w)&=&\mp \sum_{m>0}\frac{a_m}{[m]_q}
q^{\mp \frac{m}{2}}w^{-m}.
\end{eqnarray}

\subsection{Ground state}

In this section we give the free field realization
of the ground state $|B\rangle_\epsilon$ which satisfies
\begin{eqnarray}
T_\epsilon(z)|B\rangle_\epsilon=|B\rangle_\epsilon.\nonumber
\end{eqnarray}

\begin{thm}~~~
The free field realizations 
of the ground states $|B\rangle_\epsilon$
are given by
\begin{eqnarray}
|B\rangle_\epsilon
=e^{F(\epsilon)}|\Lambda_1 \rangle,~~
(\epsilon=\pm,0).
\end{eqnarray}
Here we have set
\begin{eqnarray}
F(\epsilon)
&=&-\frac{1}{2}\sum_{m>0}\frac{mq^{8m}}{
[2m]_q-(-1)^m[m]_q}a_{-m}^2\\
&+&\sum_{m>0}
\left\{\theta_m\left(\frac{
(q^{\frac{m}{2}}-q^{-\frac{m}{2}}-(\sqrt{-1})^m)q^{4m}}{[2m]_q-(-1)^m[m]_q}\right)
-\frac{(\epsilon \sqrt{-1})^{m} q^{3m}}
{[2m]_q-(-1)^m[m]_q}\right\} a_{-m}-Q,\nonumber
\end{eqnarray}
where  
$$\theta_m(x)=\left\{\begin{array}{cc}
x,&m:{\rm even}\\
0,&m:{\rm odd}
\end{array}\right..
$$
\end{thm}
Multiplying the type-II vertex operators 
$\widetilde{\Psi}_\mu^*(\xi)$
to the ground state $|B\rangle_\epsilon$,
we get the diagonalization of 
the transfer matrix $T_\epsilon(z)$
on the space of state.

~\\
{\it Proof.}~~~
Let us multiply the vertex operator $\Phi_j(z)$
to the relation $T_\epsilon(z)|B\rangle_\epsilon=
|B\rangle_\epsilon$ from the left and use
the inversion relation (\ref{eqn:VOinv}).
Then we have
\begin{eqnarray}
K_\epsilon(z)_j^j \Phi_j(z)|B\rangle_\epsilon=
\Phi_j(z^{-1})|B\rangle_\epsilon,~~~(j, \epsilon=\pm,0).
\label{def:ground2}
\end{eqnarray}
We would like to calculate
the action of the vertex operator $\Phi_j(z)$
on the vector $|B\rangle_\epsilon$.
Using the normal orderings
\begin{eqnarray}
\Phi_-(z)x^-(x^4w)&=&:\Phi_-(z)x^-(x^4w):
\frac{-1}{z(1-qw/z)},~~~(|qw|<|z|),\\
x^-(x^4w)\Phi_-(z)&=&
:x^-(x^4w)\Phi_-(z):
\frac{1}{w(1-qz/w)},~~~(|qz|<|w|),\\
x^-(w_1)x^-(w_2)&=&:
x^-(w_1)x^-(w_2):\frac{w_1(1-q^2w_2/w_1)(1-w_2/w_1)}{
(1+qw_2/w_1)},~(|w_2|<|w_1|),\nonumber\\
\end{eqnarray}
we have following realizations of the vertex operators 
$\Phi_j(z)$.
\begin{eqnarray}
\Phi_-(z)&=&\frac{1}{\epsilon(q)}
e^{P(z)}e^{Q(z)}e^Q(-zq^4)^{P+\frac{1}{2}},
\\
\Phi_0(z)&=&\oint_{C_1} \frac{dw}{2\pi \sqrt{-1}w}
\frac{(q^2-1)}{q^4z (1-qw/z)(1-qz/w)}:\Phi_-(z)x^-(q^4w):
,\\
\Phi_+(z)&=&\oint \oint_{C_2} 
\frac{dw_1}{2\pi \sqrt{-1}w_1}
\frac{dw_2}{2\pi \sqrt{-1}w_2}
q^{-\frac{5}{2}}
(1-q^2)^2\nonumber\\
&\times&
\frac{(1-w_1/w_2)(1-w_2/w_1)}
{(1+qw_1/w_2)(1+qw_2/w_1)}
\frac{z^{-2}\{(q+q^{-1})z-(w_1+w_2)\}}{
\displaystyle
\prod_{j=1}^2(1-qw_j/z)(1-qz/w_j)}\nonumber\\
&\times&
:\Phi_-(z)x^-(q^4w_1)x^-(q^4w_2):.
\end{eqnarray}
The integration contour $C_1$
encircles $w=0,qz$ but not $w=q^{-1}z$.
The integration contour $C_2$
encircles $w_1=0,qz,qw_2$ 
but not $w_1=q^{-1}z,q^{-1}w_2$, and
encircles $w_2=0,qz,qw_1$ 
but not $w_2=q^{-1}z,q^{-1}w_1$.

The actions of the basic operators 
$e^{S^-(w)}, e^{Q(z)}$
on the vector $|B\rangle_\epsilon$
have the following formulae.
\begin{eqnarray}
e^{Q(z)}|B\rangle_\epsilon=\varphi_\epsilon(z^{-1})
e^{P(z^{-1})}|B\rangle_\epsilon,\\
e^{S^-(q^4w)}|B\rangle_\epsilon=
g_\epsilon (w)e^{R^-(q^4/w)}|B\rangle_\epsilon,
\end{eqnarray}
where $\varphi_\epsilon(z)$ is given in 
(\ref{def:varphi})
and
$g_\epsilon(w)$ is given by
\begin{eqnarray}
g_\epsilon(w)=\left\{
\begin{array}{cc}
(1-w^{-2}),&~~~(\epsilon=0)\\
(1-w^{-2})(1\mp \sqrt{-1}q^{-\frac{1}{2}}w^{-1}),
&~~~(\epsilon=\pm)
\end{array}\right..
\end{eqnarray}
We have the action of the vertex operators
$\Phi_j(z)$ on the vector $|B\rangle_\epsilon$
as following.
\begin{eqnarray}
\Phi_-(z)|B\rangle_\epsilon
&=&\frac{1}{\epsilon(q)} 
\varphi_\epsilon(z^{-1})
e^{P(z)+P(z^{-1})}e^{Q}
e^{F(\epsilon)}|\Lambda_1\rangle,
\label{eqn:action1}\\
\Phi_0(z)|B\rangle_\epsilon
&=&(1-q^2)\varphi_\epsilon(z^{-1})
\oint_{\widetilde{C}_1}\frac{dw}{2\pi\sqrt{-1} w}
\frac{z^{-1} w g_\epsilon(w)}{
(1-qw/z)(1-qz/w)(1-q/zw)}\nonumber\\
&\times&
e^{P(z)+P(z^{-1})+R^-(q^4w)+R^-(q^4/w)}
e^{F(\epsilon)}|\Lambda_1 \rangle,
\label{eqn:action2}
\\
\Phi_+(z)|B\rangle_\epsilon&=&
(1-q^2)^2 q^{-\frac{5}{2}} \epsilon(q)
\varphi_\epsilon(z^{-1})
\oint \oint_{\widetilde{C}_2} 
\frac{dw_1}{2\pi\sqrt{-1}w_1}
\frac{dw_2}{2\pi\sqrt{-1}w_2}\nonumber\\
&\times&
\frac{(1-w_1/w_2)(1-w_2/w_1)(1-1/w_1w_2)(1-q^2/w_1w_2)}{
(1+qw_1/w_2)(1+qw_2/w_1)(1+q/w_1w_2)} \prod_{j=1}^2
w_j g_\epsilon(w_j)\nonumber\\
&\times&
\frac{z^{-2} \{(q+q^{-1})z-(w_1+w_2)\}}{
\displaystyle
\prod_{j=1}^2(1-qw_j/z)(1-qz/w_j)(1-q/zw_j)}
\label{eqn:action3}
\\
&\times&
e^{P(z)+P(z^{-1})+R^-(q^4w_1)+R^-(q^4/w_1)
+R^-(q^4w_2)+R^-(q^4/w_2)}e^{-Q}e^{F(\epsilon)}
|\Lambda_1\rangle.\nonumber
\end{eqnarray}
The integration contour $\widetilde{C}_1$
encircles $w=0,qz,qz^{-1}$ but not $w=q^{-1}z$.
The integration contour $\widetilde{C}_2$
encircles $w_1=0,qz,qz^{-1}, qw_2, qw_2^{-1}$ 
but not $w_1=q^{-1}z,q^{-1}w_2$,
and encircles $w_2=0,qz,qz^{-1}, qw_1, qw_1^{-1}$ 
but not $w_2=q^{-1}z,q^{-1}w_1$.

For simplicity we summarize the case $\epsilon=\pm$.
The relation (\ref{def:ground2}) is equivalent to
the following three relations.
\begin{eqnarray}
\varphi_\pm (z) \Phi_-(z)|B\rangle_\pm
&=&
\varphi_\pm (z^{-1}) \Phi_-(z^{-1})|B\rangle_\pm,
\label{def:ground3-1}
\\
\varphi_\pm(z) (1\mp \sqrt{-1}q^{-\frac{3}{2}}z) \Phi_0(z)|B\rangle_\pm
&=&
\varphi_\pm(z^{-1}) 
(1\mp \sqrt{-1} q^{-\frac{3}{2}}z^{-1}) 
\Phi_0(z^{-1})|B\rangle_\pm,
\label{def:ground3-2}\\
\varphi_\pm (z) z \Phi_+(z)|B\rangle_\pm
&=&
\varphi_\pm (z^{-1}) z^{-1} \Phi_+(z^{-1})|B\rangle_\pm,
\label{def:ground3-3}
\end{eqnarray}
$\bullet$~The relation (\ref{def:ground3-1}).
Using formula (\ref{eqn:action1}), we have
\begin{eqnarray}
(LHS)=\frac{1}{\epsilon(q)}\varphi_\pm(z)
\varphi_\pm(z^{-1})e^{P(z)+P(z^{-1})}e^{Q}
e^{F(\epsilon)}|\Lambda_1 \rangle
=(RHS).
\end{eqnarray}
$\bullet$~The relation (\ref{def:ground3-2}).
Using formula (\ref{eqn:action2}), we have
\begin{eqnarray}
&&(LHS)-(RHS)
=(1-q^2) \varphi_\pm(z)\varphi_\pm(z^{-1}) (z^{-1}-z)\\
&\times&\oint_{\widehat{C}_1} \frac{dw}{2\pi \sqrt{-1}w}
I_1(w) \frac{e^{P(z)+P(z^{-1})+R^-(q^4w)+R^-(q^4w^{-1})}
}{(1-qw/z)(1-qz/w)(1-q/wz)(1-qwz)}
e^{F(\epsilon)}|\Lambda_1 \rangle,\nonumber
\end{eqnarray}
where we have set
$
I_1(w)=(w-w^{-1})
(1\mp \sqrt{-1}q^{-\frac{1}{2}}w)
(1\mp \sqrt{-1}q^{-\frac{1}{2}}w^{-1})$.
Here the integration contour $\widehat{C}_1$
encircles $w=0,qz,qz^{-1}$ but not $w=q^{-1}z, q^{-1}z^{-1}$.
The integration contour $\widehat{C}_1$ and
the integrand
$$\frac{e^{P(z)+P(z^{-1})+R^-(q^4w)+R^-(q^4w^{-1})}
}{(1-qw/z)(1-qz/w)(1-q/wz)(1-qwz)}$$
are invariant
under $w \to w^{-1}$.
Hence the relation $I_1(w)+I_1(w^{-1})=0$ ensures
$(LHS)-(RHS)=0$.\\
$\bullet$~The relation (\ref{def:ground3-3}).
Using formula (\ref{eqn:action3}), we have
\begin{eqnarray}
(LHS)-(RHS)&=&
q^{-\frac{5}{2}}(1-q^2)^2 \epsilon(q)
\varphi_\pm(z)
\varphi_\pm(z^{-1})
\oint \oint_{\widehat{C}_2}
\frac{dw_1}{2\pi\sqrt{-1}w_1}
\frac{dw_2}{2\pi\sqrt{-1}w_2}\nonumber\\
&\times&
\frac{(1-w_1/w_2)(1-w_2/w_1)(1-w_1w_2)(1-1/w_1w_2)}{
(1+qw_1/w_2)(1+qw_2/w_1)(1+qw_1w_2)(1+q/w_1w_2)} 
\nonumber\\
&\times&
\frac{q^2 (z-z^{-1})I_2(w_1,w_2)}{
\displaystyle \prod_{j=1}^2
(1-qz/w_j)(1-qw_j/z)(1-qzw_j)(1-q/zw_j)}
\\
&\times&
e^{P(z)+P(z^{-1})+R^-(q^4w_1)+R^-(q^4/w_1)
+R^-(q^4w_2)+R^-(q^4/w_2)}e^{-Q}e^{F(\epsilon)}
|\Lambda_1 \rangle,\nonumber
\end{eqnarray}
where we have set
\begin{eqnarray}
I_2(w_1,w_2)&=&
\{-(q+q^{-1})(z+z^{-1})w_1w_2+(w_1+w_2)(w_1w_2+1)\}
\nonumber\\
&\times&
\frac{(1+qw_1w_2)}{(1-w_1w_2)}(1-q^2/w_1w_2)
\prod_{j=1}^2 (w_j-w_j^{-1})
(1\mp \sqrt{-1}q^{-\frac{1}{2}}w_j^{-1}).
\end{eqnarray}
Here 
the integration contour $\widehat{C}_2$
encircles $w_1=0,qz,qz^{-1}$, 
$qw_2, qw_2^{-1}$ 
but not
$w_1=q^{-1}z, q^{-1}z^{-1}$,
$q^{-1}w_2, q^{-1}w_2^{-1}$,
and encircles
$w_2=0,qz,qz^{-1}$, $qw_1, qw_1^{-1}$ 
but not
$w_2=q^{-1}z, q^{-1}z^{-1}$,
$q^{-1}w_1, q^{-1}w_1^{-1}$.
The integration contour $\widehat{C}_2$
and
the integrand
\begin{eqnarray}
&&\frac{(1-w_1/w_2)(1-w_2/w_1)(1-w_1w_2)(1-1/w_1w_2)}{
(1+qw_1/w_2)(1+qw_2/w_1)(1+qw_1w_2)(1+q/w_1w_2)} 
\nonumber\\
&\times&
\frac{
e^{P(z)+P(z^{-1})+R^-(q^4w_1)+R^-(q^4/w_1)
+R^-(q^4w_2)+R^-(q^4/w_2)}
}{\displaystyle
\prod_{j=1}^2
(1-qz/w_j)(1-qw_j/z)(1-qzw_j)(1-q/zw_j)}\nonumber
\end{eqnarray}
are invariant under 
$(w_1,w_2)\to(w_1^{-1},w_2),(w_1,w_2^{-1}),
(w_1^{-1},w_2^{-1})$.
Hence the relation
$$
I_2(w_1,w_2)+I_2(w_1^{-1},w_2)
+I_2(w_1,w_2^{-1})+I_2(w_1^{-1},w_2^{-1})=0
$$
ensures $(LHS)-(RHS)=0$.\\
Q.E.D.

\subsection{Dual ground state}

In this section we give the free field realizations
of the dual ground state 
$~_\epsilon\langle B| \in V(\Lambda_1)^*, (\epsilon=\pm,0)$,
which satisfies
\begin{eqnarray}
~_\epsilon\langle B|T_\epsilon(z)=
~_\epsilon\langle B|,~~~(\epsilon=\pm,0).
\label{def:dualground}
\end{eqnarray}
The dual integrable highest weight representation 
$V(\Lambda_1)^*$
of $U_q(A_2^{(2)})$ is realized by
\begin{eqnarray}
V(\Lambda_1)^*=\langle \Lambda_1|
\oplus_{n \in {\mathbb Z}}e^{nQ}
{\mathbb C}[a_{1},a_{2},\cdots],~~
\langle \Lambda_1|=\langle 0|
e^{-\frac{Q}{2}}.
\end{eqnarray}
The vacuum vector $\langle 0|$ is characterized by
\begin{eqnarray}
\langle 0|a_{-m}=0,~(m>0),~~\langle 0|P=0. 
\end{eqnarray}

\begin{thm}~~
The free field realizations 
of the dual ground states $~_\epsilon\langle B|$
are given by
\begin{eqnarray}
~_\epsilon \langle B|
=\langle \Lambda_1|e^{G(\epsilon)},~~
(\epsilon=\pm,0).
\end{eqnarray}
Here we have set
\begin{eqnarray}
G(\epsilon)
&=&-\frac{1}{2}\sum_{m>0}\frac{mq^{-2m}}{
[2m]_q-(-1)^m[m]_q}a_{m}^2\\
&+&\sum_{m>0}
\left\{-\theta_m\left(\frac{
(q^{\frac{m}{2}}-q^{-\frac{m}{2}}-(\sqrt{-1})^m)q^{-m}}{
[2m]_q-(-1)^m[m]_q}\right)
+\frac{(-\epsilon \sqrt{-1})^{m} q^{-2m}}
{[2m]_q-(-1)^m[m]_q}\right\} a_{m}-Q.\nonumber
\end{eqnarray}
\end{thm}
The proof of (\ref{def:dualground})
is given as the same way as those of (\ref{def:ground}).
The following relations  are useful for proof.
\begin{eqnarray}
~_\epsilon\langle B|e^{P(-q^{-3}z^{-1})}&=&
\varphi_{\epsilon}(z^{-1})
~_\epsilon\langle B|e^{Q(-q^{-3}z)},\\
~_\epsilon\langle B|e^{R^-(qw)}&=&
g_\epsilon^*(w)
~_\epsilon\langle B|
e^{S^-(qw^{-1})},
\end{eqnarray}
where $\varphi_\epsilon(z)$ is given in 
(\ref{def:varphi})
and
$g_\epsilon^*(w)$ is given by
\begin{eqnarray}
g_\epsilon^*(w)=\left\{
\begin{array}{cc}
(1-w^2),&~~~(\epsilon=0)\\
(1-w^2)(1\pm \sqrt{-1}q^{-\frac{1}{2}}w),
&~~~(\epsilon=\pm)
\end{array}\right..
\end{eqnarray}

\section*{Acknowledgments}

This work is supported by the Grant-in-Aid for
Scientific Research {\bf C} (21540228)
from Japan Society for Promotion of Science.


\begin{thebibliography}{0}    


\bibitem{JM}M.Jimbo and T.Miwa,
{\it Algebraic Analysis of Solvable Lattice Models},
(CBMS Regional Conference Series {\bf 85},
American Mathematical Society 1995).
\bibitem{JKKKM}M.Jimbo, R.Kedem, T.Kojima, H.Konno and T.Miwa,
{\it Nucl.Phys.}{\bf B441} 437 (1995).
\bibitem{Sklyanin}E.K.Sklyanin, 
{\it J.Phys.}{\bf A21} 2375 (1988).
\bibitem{IK}A.G.Izergin and V.E.Korepin,
{\it Commun.Math.Phys.}{\bf 79} 303 (1981).
\bibitem{BFKZ}M.T.Batchelor, V.Fridkin, A.Kuniba and Y.K.Zhou,
{\it Phys.Lett.}{\bf B376} 266 (1996).
\bibitem{MN}L.Mezincescu and R.I.Nepomechie,
{\it Int.J.Mod.Phys.}{\bf A6} 5231 (1991).
\bibitem{YZ}W.-L.Yang and Y.-Z.Zhang, 
{\it Nucl.Phys.}{\bf B596} 495 (2001).
\bibitem{FK}H.Furutsu and T.Kojima,
{\it J.Math.Phys.}{\bf 41} 4413 (2000).
\bibitem{MW}T.Miwa and R.Weston,
{\it Nucl.Phys.}{\bf B486} 517 (1997).
\bibitem{K}T.Kojima,
{\it J.Math.Phys.}{\bf 52} 01351 (2011).
\bibitem{Bax}R.J.Baxter, {\it Exactly Solved Models in
Statistical Mechanics}, 
(Academic Press, London, 1982).
\bibitem{KKMMNN}S.J.Kang, M.Kashiwara,
K.Misra, T.Miwa, T.Nakashima and A.Nakayashiki,
{\it Int.J.Mod.Phys.}{\bf A7} 449 (1992).
\bibitem{VR}V.I.Vichirko and N.Yu.Reshetikhin,
{\it Theor.Math.Phys.}{\bf 56}
805 (1983).
\bibitem{Matsuno}Y.Matsuno, Thesis (Kyoto University) 1997.
\bibitem{Jing}N.H.Jing,
{\it Invent.Math.}{\bf 102} 663 (1990).
\bibitem{JingMisra}N.H.Jing and K.C.Misra,
{\it Trans.Amer.Math.Soc.} {\bf 351} 1663 (1999).
\bibitem{HYZ}
B.-Y.Hou, W.-L.Yang and Y.-Z.Zhang,
{\it Nucl.Phys.} {\bf B556} 485 (1999).

\end{thebibliography}
\end{document}